\documentclass[twocolumn,prb,amssymb]{revtex4}

\usepackage[dvips]{graphicx}
\usepackage{amssymb}
\usepackage{amsmath}

\begin{document}

\title{Sliding Friction at a Rubber/Brush Interface}

\author{Lionel Bureau,
Liliane L\'eger
}

\affiliation{Laboratoire de Physique des Fluides Organis\'es, UMR
7125 Coll\`ege de France-CNRS, 11 place Marcelin Berthelot, 75231
Paris Cedex 05, France}

\begin{abstract}We study the friction of a poly(dimethylsiloxane) (PDMS)
rubber network sliding, at low velocity, on a substrate on which
PDMS chains are end-tethered. We thus clearly evidence the
contribution to friction of the pull-out mechanism of chain-ends
that penetrate into the network. This interfacial dissipative
process is systematically investigated by probing the velocity
dependence of the friction stress and its variations with the
grafting density and molecular weight of the tethered chains. This allows
us to confirm semi-quantitatively the picture of arm retraction relaxation of the grafted chains
proposed in models of slippage at a network/brush interface.
\end{abstract}

\maketitle

\section{Introduction}

The mechanical properties of an interface between polymers
strongly depend on the ability to form entanglements by
interdiffusion of chains between the bodies in contact. Extraction
of these bridging chains then plays a major role in energy
dissipation during the rupture of the interface. This pull-out
mechanism, which is relevant in adhesion
\cite{Lrev1,Deruelle,Creton}, seems to be also most important in
friction. Indeed, different experimental studies of the shear
properties of ``model'' interfaces, formed between an entangled or
crosslinked sample and a solid surface on which chains are
attached or adsorbed, have shown that friction is influenced by
the presence of connecting macromolecules at the interface
\cite{Lrev2}. This is for instance the case in problems of
slippage of polymer melts flowing along a
surface\cite{Lrev1,Lrev2,Dubbeldam,Durliat}: Durliat {\it et al.}
\cite{Durliat} thus evidenced the role of surface-grafted chains
on the transition between high and low interfacial friction when a
poly(dimethylsiloxane) (PDMS) melt flows on a brush or
pseudo-brush of the same polymer. L\'eger {\it et al.} \cite{SBR}
also showed that for melts of Styrene-Butadiene Rubber (SBR), the
entanglement-disentanglement mechanism between bulk and surface
macromolecules could give rise to a ``stick-slip" behaviour.

In these examples, the overall response of the system is the combined result of both the
surface chains dynamics and the bulk chains reptation.
In order to probe the effect of grafted molecules in the simpler
case where the bulk is a permanent network, other authors
investigated the velocity-dependent frictional response of an
elastomer sliding on a brush: Brown \cite{Brown} first pointed out that
tethered chains could either enhance or lower the friction,
depending on their areal density. Casoli {\it et al.} \cite{Casoli} further studied
the behaviour of such rubber/brush interfaces at high sliding
velocities
and showed that the presence of the surface-anchored chains could
lead to strong departure from the expected linear relation between
friction force and velocity.

A central question for understanding these polymer friction problems is that of the dynamics of  penetration
of an end-tethered chain into a network. This has been the subject of theoretical \cite{Mcleish} and numerical \cite{Deutsch}
works in the case of a static interface,
which show that after
an initial stage of rapid penetration --- on time scales on the order
of the Rouse time ($\tau_R$) of the chain --- the dynamics is
controlled by arm retraction of the grafted chain, which results in
slow relaxations (logarithmic in time). This arm retraction mechanism is also expected to
control friction when a highly entangled polymer melt slips on a grafted surface
\cite{Rubinstein, Ajdari,Deutsch2}. Different regimes for the velocity dependence
of the friction force are thus predicted, depending on the respective values of the advection time, $D_e/V$ (where
$D_e$ is the mesh size of the network formed by the melt), and
the relaxation time
$\tau_{arm}$ of all or part of the grafted chain.

Though a rather good agreement has been obtained between experimental results on melts flows and the
above theoretical predictions for the threshold velocities between the different regimes \cite{Lrev2},
there is up to now no quantitative test of the model concerning the velocity dependence of the friction force.
Indeed, the picture given in the model for frictional dissipation by pull-out--- though such a mechanism must actually take
place at the interface --- cannot be validated from the existing experimental data, since no systematic investigation
of the friction force has been done on systems where the molecular control parameters are varied.

In this paper, we present an experimental study of friction of a rubber network sliding on a brush, at low slip velocities
such that the grafted chains are able to penetrate in the elastomer. We first briefly describe the experimental
techniques used for samples preparation and for friction force measurements. The different results obtained, which depend on
grafting density and molecular weight of the tethered chains, are then presented and discussed in the framework of the
existing friction model cited above.

\section{Experiments}

\begin{figure}[htbp]
$$
\includegraphics[scale=0.8]{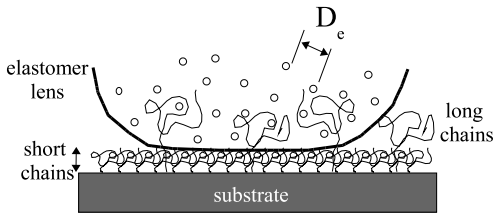}
$$
\caption{Sketch of the contact configuration involved in the experiments: a lens of PDMS elastomer of
mean mesh size $D_e$ is in contact with a bimodal brush made of densely grafted short chains and long connectors which
can penetrate into the network.}
\label{fig1}
\end{figure}

Our experiments involve the contact between a lens of
poly(dimethylsiloxane) (PDMS) elastomer and a ``bimodal'' brush
--- {\it i.e.} made of chains of two different length --- grafted on a
flat substrate (figure \ref{fig1}).

\subsection{Samples preparation}

Elastomers are obtained using the following method \cite{Deruelle}:
chains of $\alpha$,$\omega$-vinyl-terminated PDMS (fractionated in
the laboratory from commercial oils (Rhodia)) are crosslinked by hydrosilation of the vinyl
ends with the Si-H groups of a
tetrafunctional crosslinker, in the presence of a Pt complex catalyst. The ratio of hydride to vinyl
functions is adjusted in order to optimize the connectivity of
the network \cite{Deruelle}, and the crosslinking is made under dry
nitrogen. Flat/convex lenses of elastomer are obtained by putting droplets of
the unreacted melt/crosslinker mixture on a non-wetting surface \cite{Chaudhury}.
After reaction at 110$^{\circ}$C for 12h, the
networks formed are washed for ten days in a solution of toluene
containing a dodecanethiol, in order to extract the unreacted chains and
to inhibit the catalyst. We thus prepared two series of PDMS
elastomers made from chains of molecular weight $M_w=9$ or 23
kg.mol$^{-1}$ ({\it i.e.} roughly once and twice the critical
weight for entanglements), of respective polydispersity $I=$1.14
and 1.17.

The bimodal brushes are prepared as follows \cite{Folkers}: a short (4
siloxane monomers) SiH-terminated monochlorosilane is first
grafted on the silica layer of a silicon wafer. A melt of PDMS
containing a mixture of short ($M_w$=5 kg.mol$^{-1}$, $I$=1.16)
and long ($M_w$=114, 89, 58, 35 or 27 kg.mol$^{-1}$, resp. $I$=1.25, 1.28, 1.60, 1.33,
1.11) chains is then spread on this H-terminated sublayer. The
chains in the melt are all
$\alpha$-vinyl,$\omega$-methyl-terminated, and grafting occurs by
hydrosilation of the vinyl ends at 110$^{\circ}$C for 12h. After
rinsing by sonication in a solution of toluene and thiol, we measure by
ellipsometry the dry thickness $h$ of the grafted layers, and
deduce from this the chains areal density $\Sigma$, which reads
$\Sigma=h/(Za^3)$, where $Z$ is the polymerization index of the
grafted chains and $a\simeq$0.5~nm the size of a monomer \cite{Deruelle}. \\
(i) Grafting
from a melt of short 5 kg.mol$^{-1}$ chains leads to a dry thickness
$h_{short}=3.6\pm 0.3$~nm, {\it i.e.} $\Sigma_{short}=0.41\pm 0.03$
nm$^{-2}$. This thickness is comparable to the size of the
unperturbed chains $R_0\sim a\sqrt{Z}\simeq$4 nm, and corresponds
to a density $\Sigma_{short}\gg 1/R_0^2$. We thus obtain brushes
of short chains which can be pictured as a dense layer of
overlapping gaussian coils. The role of this layer is to prevent
direct contact between the elastomer and the silane sublayer or the silica surface.\\
(ii) Choosing the weight ratio of short to long chains in the
grafting mixture then allows to obtain bimodal brushes of a given
density of long chains \cite{Durliat},
$\Sigma=(h-h_{short})/(Za^3)$, which can be adjusted between 0 and
0.08 nm$^{-2}$ for the different molecular weights used here (the
total thickness $h$ ranges from $h_{short}$ to $h\simeq 15$ nm for
the highest $M_w$). In the following, $\Sigma$ always refers to
the areal density of the long ``connectors'', with $\Sigma>
1/R_0^2$ except for the two lowest densities which are in the
``mushroom'' regime of non overlapping chains.

We systematically prepare
brushes of various $\Sigma$, along with a
reference brush with $\Sigma =0$, from the same
wafer, {\it i.e.} from the same H-terminated silane layer.

\subsection{Force measurement setup}

\begin{figure}[htbp]
$$
\includegraphics[scale=0.8]{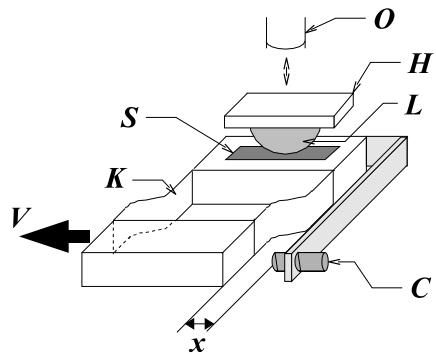}
$$
\caption{Experimental setup: the lens (L), adhered on a glass plate (H), is in contact with the substrate (S). (S) is
moved through a spring (K) which is driven at constant velocity. A capacitive sensor (C) measures the spring bending.
The contact area is monitored optically by means of a long working distance objective (O).}
\label{fig2}
\end{figure}

Friction measurements are performed on a setup designed on purpose for
this study (figure \ref{fig2}): a lens of elastomer, maintained by
a horizontal glass plate, is brought in contact with a
brush-bearing substrate. This substrate is fixed at the free end
of a double cantilever spring (stiffness 400 N.m$^{-1}$), the
second end of which is driven at constant velocity
3~nm.s$^{-1}<V<$330~$\mu$m.s$^{-1}$. A capacitive displacement
gauge measures the spring bending, which allows us to access the
friction force $F$ in the range 50 $\mu$N--50 mN. The contact area
$S$ between the lens and the flat is monitored optically. The
normal force is not measured in this experiment but the
sphere/flat indentation depth is fixed instead, which amounts to
work at given contact diameter, in the range 200--400 $\mu$m.

\section{Results}
\label{results}

\begin{figure}[htbp]
$$
\includegraphics[scale=1]{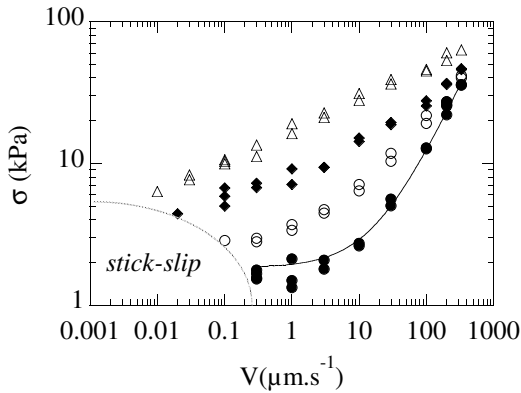}
$$
\caption{Friction stress $\sigma$ {\it vs} sliding velocity $V$ (both scales are logarithmic), for connectors of $M_w=114$
kg.mol$^{-1}$ and elastomer of $M_w$=23 kg.mol$^{-1}$. ($\bullet$): $\Sigma=0$; ($\circ$): $\Sigma=0.0016$ nm$^{-2}$; ($\blacklozenge$): $\Sigma=0.0076$ nm$^{-2}$;
($\triangle$): $\Sigma=0.021$ nm$^{-2}$. The solid line is a fit of the $\Sigma=0$ data with $\sigma=\sigma_0+kV$,
$\sigma_0=1.8$ kPa, $k=10^8$ Pa.s.m$^{-1}$. Stick-slip is systematically observed at low velocities.}
\label{fig3}
\end{figure}

Typical results obtained in steady sliding are presented in figure
\ref{fig3}, where the mean friction stress $\sigma=F/S$ is plotted
against sliding velocity for various grafting densities of
connectors. We have checked that $\sigma$ does not depend  on the contact size $S$ in the experimentally
accessible range.

\subsection{Friction on short chains brushes}

At $\Sigma=0$, {\it i.e.} when sliding on the dense brush of short chains only, the stress increases linearly with
velocity after a low-velocity plateau. As seen in figure \ref{fig3}, the data are well fitted by
$\sigma=\sigma_0+k V$. The plateau value $\sigma_0$ can vary from 2 to 10 MPa from sample to sample, but the coefficient $k$
takes a constant value $k=10^8$ Pa.s.m$^{-1}$.

The steady sliding regime does not span the whole range of
velocities: at $V$ lower than a
critical value $V_c$, we observe stick-slip oscillations, {\it i.e.} unstable sliding, down to the lowest accessible
velocity. The bifurcation from stable to unstable sliding is continuous, without hysteresis: the force signal smoothly goes from small
sine-shaped
oscillations at $V\sim V_c$ to large amplitude triangular oscillations at $V\ll V_c$. The critical velocity $V_c$ can vary
from 0.1 to 3 $\mu$m.s$^{-1}$ from sample to sample.

The scattering on $V_c$ and $\sigma_0$ mentioned above will be discussed
in section \ref{discussion}.

\subsection{Role of connectors}

Adding long connectors, whatever their molecular weight, to the short chains brushes has different effects:

(i) The friction stress depends non-linearly on velocity, and tends to grow
as a weak power law of $V$ at high $\Sigma$ (figure \ref{fig3}).

\begin{figure}[htbp]
$$
\includegraphics[scale=1]{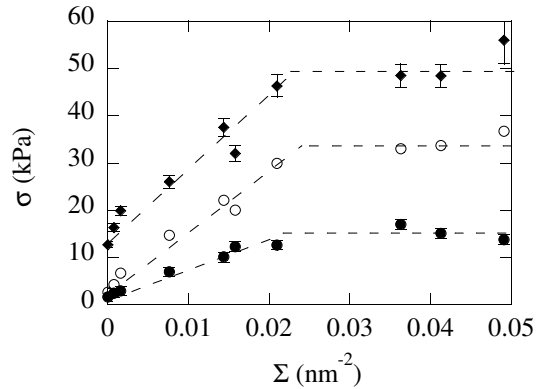}
$$
\caption{Friction stress $\sigma$ {\it vs} grafting density $\Sigma$ for connectors of $M_w=114$
kg.mol$^{-1}$ and elastomer of $M_w$=23 kg.mol$^{-1}$, at
$V=0.3$ $\mu$m.s$^{-1}$ ($\bullet$), $V=10$ $\mu$m.s$^{-1}$ ($\circ$), $V=100$ $\mu$m.s$^{-1}$ ($\blacklozenge$).
}
\label{fig4}
\end{figure}

(ii) The stress in steady-sliding {\it increases}
quasi-linearly with the grafting
density (see
figures \ref{fig3} and \ref{fig4}), up to a crossover value of the density $\Sigma_0\simeq 0.02$--0.03 nm$^{-2}$,
 above which $\sigma$ reaches a plateau, as can be seen in figure \ref{fig4}. The slope of the quasi-linear part of
$\sigma (\Sigma)$ slightly depends on the velocity at which the stress is measured, and increases, typically,
by a factor of 3 or 4 when $V$ varies over 3 decades.

\begin{figure}[htbp]
$$
\includegraphics[scale=1]{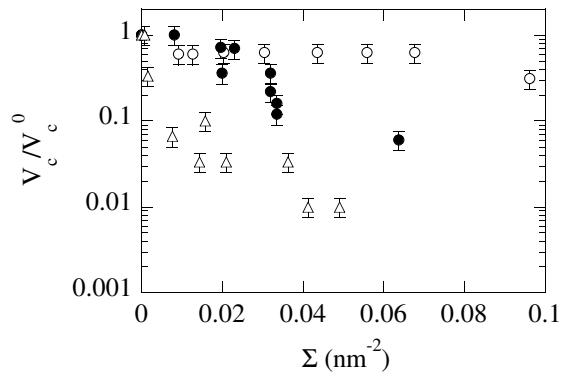}
$$
\caption{Critical velocity $V_c$ (normalized by its value at $\Sigma=0$) as a function of grafting density, for
$M_w=27$ kg.mol$^{-1}$ ($\circ$); $M_w=58$ kg.mol$^{-1}$ ($\bullet$);
$M_w=114$ kg.mol$^{-1}$ ($\triangle$).}
\label{fig5}
\end{figure}

(iii) The critical velocity $V_c$ decreases with the grafting density of connectors, as illustrated in figure
\ref{fig5}.

\subsection{Molecular weight effects}

\begin{figure}[htbp]
$$
\includegraphics[scale=1]{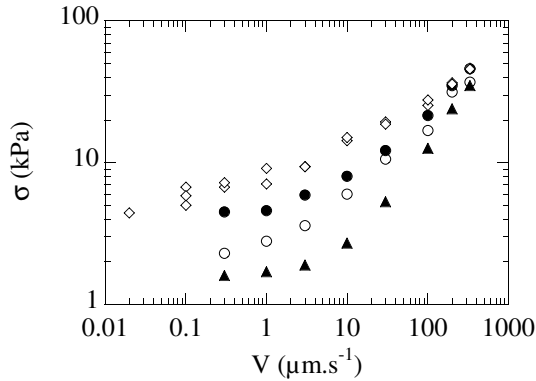}
$$
\caption{Friction stress $\sigma$ {\it vs} sliding velocity $V$, for $\Sigma=0$ ($\blacktriangle$) and for $\Sigma\simeq 0.01$
nm$^{-2}$ of different connectors.($\circ$): $M_w=27$ kg.mol$^{-1}$;  ($\bullet$): $M_w=35$ kg.mol$^{-1}$; ($\lozenge$):
$M_w=114$ kg.mol$^{-1}$.}
\label{fig6}
\end{figure}

We observe the following when varying the molecular weight of the connectors:

(i) At a given areal density $\Sigma<\Sigma_0$ ({\it i.e.} in the linear regime of $\sigma(\Sigma)$), the sliding stress $\sigma$ increases with the molecular weight of the
long chains, as illustrated in figure \ref{fig6}.

(ii) At high grafting density, the $\sigma(V)$ power law exponent grows from 0.2 to 0.5 as $M_w$
decreases from 114 to 27 kg.mol$^{-1}$.

(iii) The critical velocity $V_c$ is lower for longer connectors (figure \ref{fig5}).

Finally, we do not notice any effect of the molecular weight between crosslink points of the elastomer. All the features
described above are quantitatively identical for $M_w=9$ or 23 kg.mol$^{-1}$.

\section{Discussion}
\label{discussion}

\subsection{Origin of the stick-slip instability}

We first want to come back on the possible mechanism responsible for the stick-slip observed at low velocities.
An entanglement/disentanglement process between the grafted chains and the network cannot be invoked here, since unstable
sliding is also observed on short chains brushes which are of low $M_w$ and hence cannot entangle with the elastomer.

\begin{figure}[htbp]
$$
\includegraphics[scale=1]{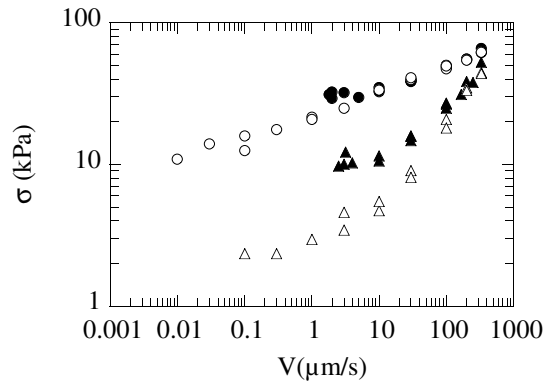}
$$
\caption{Sliding stress as a function of velocity. ($\triangle$): short chains brush $\Sigma_{short}=0.37$ nm$^{-2}$;
($\blacktriangle$): short chains brush $\Sigma_{short}=0.43$ nm$^{-2}$; ($\bullet$): $\Sigma=0.05$ nm$^{-2}$ of
$M_w=114$ kg.mol$^{-1}$ connectors and low density sublayer; ($\circ$): $\Sigma=0.05$ nm$^{-2}$ of
$M_w=114$ kg.mol$^{-1}$ connectors and high density sublayer}
\label{fig7}
\end{figure}

However, we observe that the critical velocity $V_c$ and the low velocity value $\sigma_0$ of the sliding stress
are systematically lower on short chains brushes which are more dense. This is illustrated in figure \ref{fig7}:
$\sigma_0\simeq 10$ kPa and $V_c\simeq 2$ $\mu$m.s$^{-1}$ on a brush of $\Sigma_{short}=0.37$ nm$^{-2}$, whereas
$\sigma_0\simeq 2$ kPa and $V_c\simeq 0.1$ $\mu$m.s$^{-1}$ on a brush of $\Sigma_{short}=0.43$ nm$^{-2}$. We note
moreover that both $\sigma (V)$ characteristics become similar at high velocities.

These features also hold on
bimodal brushes with the same grafting density of long chains but having short chains sublayers of different
densities
\cite{note}
(figure \ref{fig7}).

This strongly suggests that the stick-slip observed in our
experiments results from the presence of defects in the short
chains brush. Indeed, the grafting method used in this study does
not lead to the highest reachable areal densities that can be
obtained from other techniques ({\it e.g.} ``grafting from''
techniques \cite{Prucker}). These defects are probably
unreacted sites from the H-terminated silane sublayer used for
hydrosilation. These sites, may transform in silanol groups
under ambient humidity, and become adsorption sites for
the PDMS chains. This leads us to propose the following picture.
Once the contact is made between the elastomer and the grafted
layer, the short chains brush swells, which allows the elastomer
chains to access the underlying adsoption sites. When sliding, the
frictional response then depends on both the elastomer/brush and
the elastomer/defects interactions:

 At high velocities, the elastomer chains do not have time to
adsorb on defects and friction is governed by the PDMS/PDMS interactions only, which results in similar $\sigma (V)$
curves whatever $\Sigma_{short}$.

At low velocities, an
adsorption/desorption mechanism under shear can occur between the network and the pinning sites (the defects), akin to
that proposed by Schallamach for rubber friction \cite{Schallamach}. This leads to:

(i) a higher sliding stress and critical velocity $V_c$
when $\Sigma_{short}$ is lower, {\it i.e.} when
the pinning sites are more numerous,

(ii) a friction stress which {\it decreases} with velocity, which is the source of the mechanical instability observed.

The low velocity
plateau $\sigma_0$, attributable to this pinning mechanism, would then appear as a minimum at the crossover between the  velocity decreasing and increasing
regimes.

This qualitative picture is consistent with the fact that the critical velocity $V_c$ is lowered in the presence of
connectors (figure \ref{fig5}), since slow penetration of the long chains in the network is needed before accessing
the pinning sites on the substrate.

\subsection{Friction induced by grafted chains}

We now concentrate on the results obtained in steady sliding.

When the elastomer slides on a dense brush of short chains, the friction stress
obeys a relation of the form $\sigma=\sigma_0 + k V$, with $k=10^8$ Pa.s.m$^{-1}$ (figure \ref{fig3}). At this type
of interface, since the brush is dense and no entanglement can be formed, we expect the frictional response to be close
to that of an interface between two surfaces of densely packed monomers: $\sigma_{mono}=\zeta_1 V/a^2$, where
$\zeta_1$ is the monomeric friction coefficient and $a$ the monomer size. From the value of $k$ above, using $a=0.5$ nm,
we obtain $\zeta_1 =ka^2=2.5 \times 10^{-11}$ N.s.m$^{-1}$. This value is fully consistent with $\zeta_1\simeq 10^{-11}$
N.s.m$^{-1}$ deduced from self-diffusion \cite{Hervet} or viscoelastic measurements \cite{Barlow}. Partial penetration
and chain
stretching effects might
account for the fact that the value deduced from our experiments is slightly larger than the previously reported ones.

The presence of long grafted chains at the interface leads to a friction stress which is systematically higher
than on the short chains brush and which depends non-linearly on sliding velocity.
This result is due to penetration
of these chains inside the network and to their subsequent pull-out. Indeed, the characteristic time scale of our
experiments, given by $D_e/V$ (where $D_e\sim a\sqrt{N_e}\simeq 5$ nm is the mesh size of the elastomer, $N_e\simeq 100$
being the number of monomers between entanglements \cite{Hervet}), ranges from
$10^{-2}$ to $10^{-5}$ s for velocities between 0.3 and 300 $\mu$m.s$^{-1}$. This is much larger than the
Rouse
time between entanglements for PDMS ($\tau_R^e = \tau_1 N_e^2\simeq 3\times 10^{-7}$ s, with $\tau_1=3\times 10^{-11}$ s),
 and also larger than the
 longest Rouse time of the grafted chains used ($\tau_R=7\times 10^{-5}$ s for
$M_w=114$ kg.mol$^{-1}$).
We are thus in a situation where the grafted chains can always relax inside the network over a distance on the
order of or greater than the mesh size of the elastomer, {\it i.e.} the first, rapid penetration step
predicted by O'Connor {\it et al} \cite{Mcleish} or Deutsch {\it et al} \cite{Deutsch} can always occur at the interface.
During shear, this penetrated part will then allow to stretch the portion of chain confined outside the network, and thus
contribute to the friction force until being pulled out.

We have shown that, at a given velocity, the friction stress increases with the grafting density of tethered chains
before reaching a plateau. This result is in full agreement with what Casoli {\it et al.} observed for
friction on a similar
PDMS rubber/brush
system \cite{Casoli}; it is also consistent with the density dependence of the adhesion energy at such interfaces, which
exhibits a plateau or a maximum at an optimal density \cite{Tardivat}. \\
The first increase of $\sigma$ with $\Sigma$ confirms the
picture of friction enhancement by chain penetration/pull-out. It is important to note that this regime is observed at
areal densities essentially larger than the mushroom limit, which means that grafted chains overlap and can
entangle outside the elastomer. Hence, though the friction stress increases quasi-linearly with $\Sigma$, we believe that
the tethered chains cannot be considered as independent and that their dynamics is influenced by interactions between
neighbours, as stated by Casoli {\it et al.} \cite{Casoli} or as observed in the simulations of
Deutsch {\it et al.} \cite{Deutsch}. \\
The friction stress plateau is attributable to the swelling limit of the elastomer, as discussed in previous studies
on that type of interface \cite{Deruelle,Casoli}:
the number of monomers that can penetrate into the network region close to the interface is limited by the network
elasticity \cite{deGennes}.

Our results seem to contrast with those obtained by Brown \cite{Brown}
in the first study of friction at such rubber/brush interfaces. Indeed, he concluded that grafted chains had mostly
a lubricating effect. This
difference is certainly due to the substrates used in the studies: Brown grafted PDMS chains in a polystyrene layer,
whereas we used
bimodal brushes. We have checked in a control experiment that the friction stress when sliding on
a thin layer of PS is at least one order of magnitude larger than when sliding on a short chains brush of PDMS. Brown's
results then come from a progressive screening of the PS substrate while the thickness of the PDMS layer
increases. Both works are thus not in contradiction but cannot be compared directly.

Eventually, the non-linearities induced by grafted chains may have different origins:

(i) The force needed to pull the
chains out of the network may depend non-linearly on velocity.

(ii) The stress due to network deformations in the vicinity of the interface might be non-linear.

(iii) The response of the thin layer formed by the portions of chains confined outside the network
may be shear-thinning, which is suggested by the power law dependence of $\sigma$ on $V$ at high grafting densities.

The relative weight of these mechanisms is not straightforward to estimate, all the more than most of our data
have been obtained in a range of grafting density where interactions between tethered chains can certainly not
be disregarded, which renders data analysis more complex. \\
We believe nonetheless that the third mechanism, {\it i.e.} the
rheology of a thin entangled layer, dominates at high $\Sigma$:
if we estimate, for instance, that
a layer of long chains ($M_w=114$ kg.mol$^{-1}$) of thickness $h\simeq 10$ nm is sheared at a rate $\dot{\gamma}=V/h$, we find that $\dot{\gamma}$
is in the range 1--10$^4$ s$^{-1}$ for 0.01 $\mu$m.s$^{-1}$ $<V<$100 $\mu$m.s$^{-1}$. Over this range the stress
$\sigma\sim V^{0.2}$, which is consistent with the shear-thinning behaviour observed both in bulk
rheology of polymer melts \cite{Mead,Massey} (at shear rates $\dot{\gamma}>\dot{\gamma}_c\sim 1/\tau_{rep}$,
where $\tau_{rep}$ is the reptation time of the chains) and in a recent study of the shear response of PDMS melts highly
confined between rigid walls \cite{Yamada}. Now, if we evaluate an
effective viscosity $\eta_{eff}=\sigma h/V$ from our data, we find that $\eta_{eff}$ decreases from $10^4$ to 10 Pa.s
when $\dot{\gamma}$ increases from 1 to 10$^4$ s$^{-1}$, with $\eta_{eff}\sim \dot{\gamma}^{-0.8}$. For comparison,
we expect, for a
melt of chains with $M_w=114$ kg.mol$^{-1}$, the bulk viscosity to decrease from its newtonian plateau $\eta_{bulk}\simeq 10$ Pa.s
above $\dot{\gamma}>1/\tau_{rep}\simeq 10^3$ s$^{-1}$. We thus clearly see that $\eta_{eff}\gg\eta_{bulk}$, which is
consistent with the fact that in our experimental situation where chains are end-tethered, the characteristic
relaxation time should rather be
on the order of the time for arm retraction \cite{Rubinstein,Ajdari,Deutsch,Milner} (see next section), which is
much larger than the reptation time governing the bulk rheology of melts.
Both the strong increase of $\eta_{eff}\gg\eta_{bulk}$ and its power law dependence are in good agreement
with what has been observed in shear response of confined melts \cite{Yamada}. This suggests that the frictional response
 at
high grafting densities is controlled by the rheology of a thin layer of chains outside the elastomer, the behaviour of
which is strongly influenced by
confinement and tethering effects. \\
On the opposite, we can now focus on the results obtained
on the brushes with the lowest $\Sigma$, in the mushroom regime, where the third mechanism should not be
relevant. We perform a detailed analysis of these results in the next section.

\subsection{Molecular weight and chain pull-out effects}

One of the last points we want to discuss is the role of molecular weights (of the grafted chains or of the elastomer) on
friction.\\
 Let us first recall that no difference was noted between the frictional response of elastomers made from
chains of $M_w=9$ or 23 kg.mol$^{-1}$, {\it i.e.} containing $N=120$ or 310 monomers. This means that on the time scale
of the network solicitation, the
entanglements which might be trapped in the chemical network do not have time to relax,
and that the effective mesh size of the elastomer is not fixed by the total number of monomers between crosslinks but
rather by the number of monomers between entanglements $N_e\simeq 100$. In the following, we will thus consider that
the mesh size of the network is $D_e\sim a\sqrt{N_e}\simeq 5$ nm. \\
The role of grafted chains length can be understood
qualitatively as follows: the longer the chains, the more entanglements they form with the network, the higher the
friction stress. This picture might however be too naive since the penetration dynamics of the chains may
 be faster for shorter chains, which would oppose the simple entanglement effect mentioned before.

This last point deserves a more refined approach. We will now analyse the low grafting densities results on the basis
of the model originally proposed by Rubinstein {\it et al.} \cite{Rubinstein,Ajdari}, in an attempt to understand more
precisely the molecular weight and the chain pull-out effects.

We briefly recall the main features of this friction model. The authors consider the situation of an elastomer sliding
at a given velocity on a weakly grafted smooth surface, the sliding stress then being $\sigma(V)=\sigma_{bare}(V) + \Sigma F_c(V)$, where
$\sigma_{bare}$ is the contribution of the bare surface and $F_c$ the force on one grafted chain. The tethered chains
are supposed to be independent. To determine $F_c$, the assumption is made that this friction situation is analogous
to that where a single chain is pulled by its head, at constant $V$, through a network of fixed obstacles. In this latter
situation, the only possible relaxation mode of the chain is arm retraction. The relaxation time of the whole chain of $Z$
monomers is that of a thermally activated process, given by \cite{Rubinstein,Ajdari,Deutsch,Milner}
\begin{equation}
\label{tau_arm}
\tau_{arm}(Z)=\tau_1 Z^2 \exp{(\mu Z/N_e)}
\end{equation}
where $\mu$ is a constant of order 2 and $\tau_1$ is a microscopic time. As long as the advection time, $D_e/V$, is
larger than $\tau_{arm}(Z)$, {\it i.e.} as long as $V<V_1=D_e/\tau_{arm}(Z)$, the whole chain can relax in the network and
the force on its head reads
\begin{equation}
\label{F1}
F_c=F_e\frac{V}{V_1}
\end{equation}
where $F_e=kT/D_e$ ($k$ is the Boltzmann constant and $T$ the temperature).

For velocities $V>V_1$, the head of the chain is stretched in straight tube of diameter $D_e$ whereas only a tail
of $q$ monomers has time to relax and adopt a ``ball'' shape, the relaxation time of the relaxed tail being
\begin{equation}
\label{tau_arm2}
\tau_{arm}(q)=\tau_1 Z^2 \exp{[\mu q^2/(ZN_e)]}
\end{equation}
In this regime, the force is the sum of two terms:
\begin{equation}
\label{F2}
F_c=kT/D_e + [Z-q(V)]\zeta_1 V
\end{equation}
where the first term of the right-hand-side is the force on the relaxed part of the chain and the second one is the
contribution, by
Rouse friction, of the stretched part.
$\zeta_1$ is a monomeric friction coefficient. The number $q(V)$ is determined by $V\tau_{arm}(q)=D_e$.

As the velocity is increased, the ball shrinks and the stretched part grows, until $q\simeq \sqrt{ZN_e}$, {\it i.e.}
until $V\simeq V_2=D_e/(\tau_1 Z^2)$. Above this threshold, the entropic barrier for arm retraction vanishes, and
the size of the ball is then controlled by the tube length fluctuations of a chain subunit of $q$ monomers, with relaxation
time
\begin{equation}
\label{tau3}
\tau_R (q)=q^4\tau_1/N_e^2
\end{equation}
The number $q$ is thus determined by $\tau_R(q)=D_e/V$ and the force is always given by expression \ref{F2}.

The contributions to friction of the ball and of the stretched part become comparable at $V\simeq V_3=V_2(Z/N_e)$, and
for $V\gg V_3$ the friction force is $F_c\simeq Z\zeta_1 V$.

Let us estimate the values of these different velocity thresholds for the parameters of our PDMS systems. We take $\zeta_1\simeq1.5\times 10^{-11}$ N.s.m$^{-1}$
(deduced from self-diffusion measurements \cite{Hervet}), $\tau_1=\zeta_1 a^2/(3\pi^2kT)\simeq3\times 10^{-11}$ s,
$Z=1540$ for the longest connectors and $Z=365$ for the shortest ones. $V_1$ varies
from $10^{-11}$ $\mu$m.s$^{-1}$ for $Z=1540$ to about 1 $\mu$m.s$^{-1}$ for $Z=365$,
$V_2$ increases from 10 to
200 $\mu$m.s$^{-1}$, and $V_3$ from 200 to 700 $\mu$m.s$^{-1}$. We can thus consider, for all
practical purposes and in view of the uncertainty on the values of the microscopic parameters, that the first
low-velocity regime is never reached experimentally in steady-sliding. The range of velocity
covered in our study should correspond to a friction force mainly governed by expression \ref{F2}, with $q$ controlled by
$\tau_{arm}$ or $\tau_R$ depending on chain length. The friction force per grafted chain should then increase
slowly, starting  from $F_e=kT/D_e\simeq 8\times 10^{-13}$ N, as velocity is increased.

\begin{figure}[htbp]
$$
\includegraphics[scale=1]{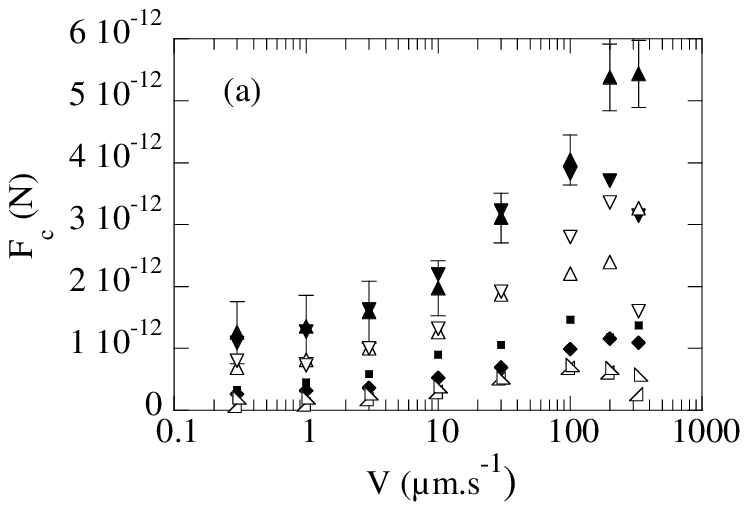}
$$
$$
\includegraphics[scale=1]{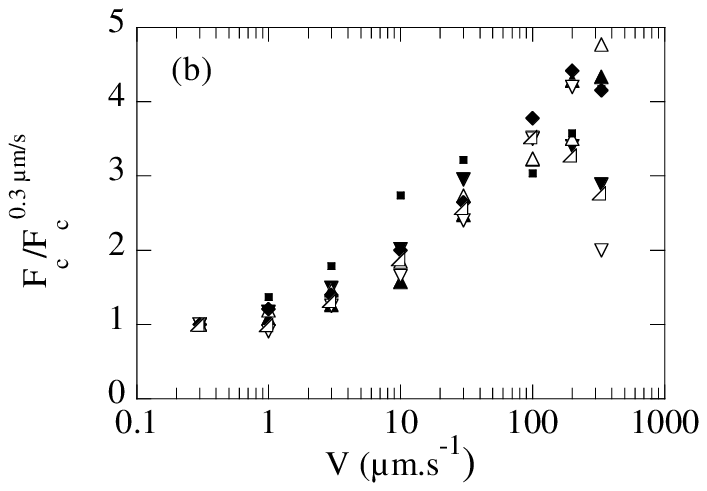}
$$
\caption{(a): friction force per grafted chain as a function of velocity. ($\blacktriangle$) and ($\blacktriangledown$):
$M_w=114$ kg.mol$^{-1}$; ($\triangle$) and ($\triangledown$):
$M_w=89$ kg.mol$^{-1}$; ($\blacksquare$) and ($\blacklozenge$):
$M_w=35$ kg.mol$^{-1}$; ($\lhd$) and ($\rhd$):
$M_w=27$ kg.mol$^{-1}$. (b) same data as in figure (a) scaled by $F_c(0.3\, \mu \text{m.s}^{-1})$.}
\label{fig8}
\end{figure}

In order to compare our results with the model predictions, we now have to extract from the data the friction force per grafted
chain. We do so by assuming that, at low grafting densities in the mushroom regime, all contributions are additive.
We consider that the total friction stress measured is $\sigma(V)=\sigma_{short}(V)+\Sigma F_c(V)$,
where $\sigma_{short}(V)$ is the friction stress measured without long connectors, and deduce $F_c$
from this. The friction force per connector thus evaluated is reported in figure \ref{fig8}, as a function of sliding velocity,
 for four different
molecular weights and two grafting densities per molecular weight.

\begin{figure}[htbp]
$$
\includegraphics[scale=1]{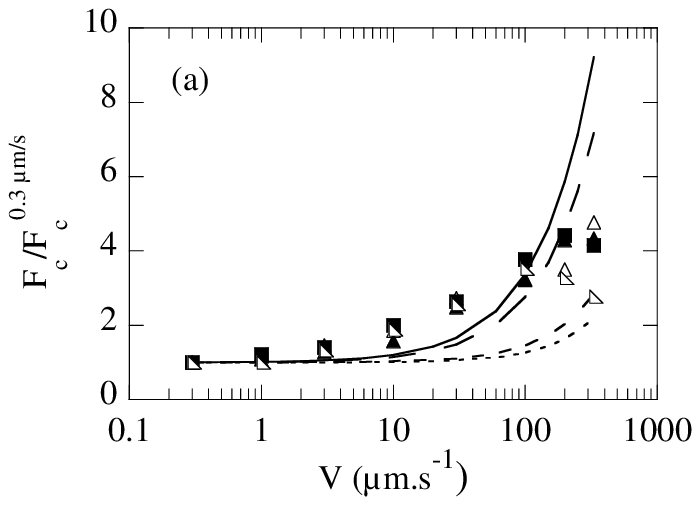}
$$
$$
\includegraphics[scale=1]{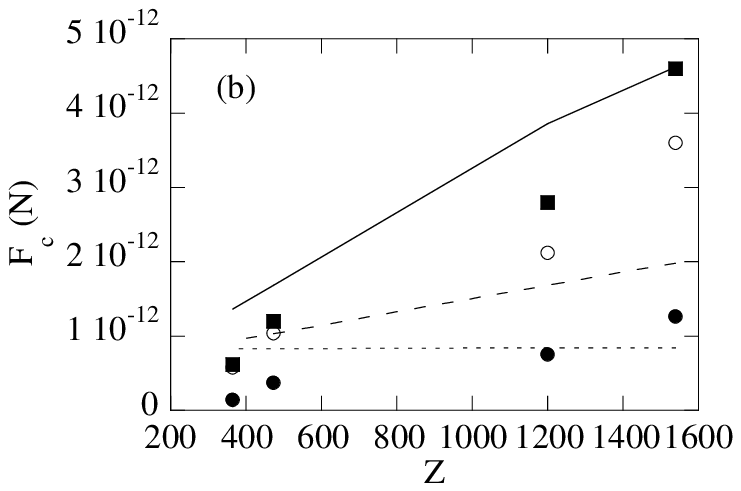}
$$
\caption{(a) Theoretical predictions for $F_c/F_e$ as a function of velocity. (---): Z=1540 monomers;
(---$\;\;$---): Z=1200;
(--$\;\;$--): Z=470; (-~-~-): Z=365. Symbols: experimental data for $F_c/F_c(0.3\, \mu \text{m.s}^{-1}))$ from figure
\ref{fig8}b. (b) Force per chain as a function of the polymerization index $Z$ of the grafted chains. Symbols:
experimental data at $V=1\, \mu$m.s$^{-1}$ ($\bullet$); $V=60\, \mu$m.s$^{-1}$ ($\circ$); $V=200\, \mu$m.s$^{-1}$
($\blacksquare$). Theoretical predictions at $V=1\,
\mu$m.s$^{-1}$ (-~-~-); $V=60\, \mu$m.s$^{-1}$ (---$\;\;$---); $V=200\, \mu$m.s$^{-1}$ (---).}
\label{fig9}
\end{figure}

It can be seen in figure \ref{fig8}a that $F_c$, as
expected, do not depend on $\Sigma$, and that the friction force decreases for shorter the connectors
(see also figure \ref{fig9}b).
In particular, the value of $F_c$ at $V=0.3$ $\mu$m.s$^{-1}$ is found to decrease from $10^{-12}$ to $10^{-13}$ N as
$M_w$ goes from 114 to 27 kg.mol$^{-1}$.
Now, if we scale the experimental curves $F_c(V)$ by the low-velocity value of $F_c$,
it appears that whatever the molecular weight of the connectors, the force per chain has the same velocity dependence
(figure \ref{fig8}b):
it increases with $V$, by a factor of 3 or 4 over three decades of velocity, and may eventually fall at the
highest velocity. Note here that this relative increase with $V$ is consistent with the slope variation of the linear part of
$\sigma(\Sigma)$ reported in section \ref{results}.

Three main features of these experimental results are in good semi-quantitative agreement with the theoretical predictions,
bearing in mind that the model is totally free of fitting parameters (we used
for the microscopic parameters $\tau_1$ and $\zeta_1$ values determined by independent techniques):

(i) the order of magnitude of the low-velocity values of the force per chain
is consistent with $F_e=kT/D_e\simeq 8\times 10^{-13}$ N,

(ii) $F_c$ increases weakly with velocity (see figure \ref{fig9}a),

(iii) $F_c$ increases quasi-linearly with chain length at high velocities (see figure \ref{fig9}b).

From this we can conclude that chain pull-out is certainly the mechanism which governs friction at low grafting
density, and that the picture of arm retraction proposed in the model accounts well for the weak velocity
dependence observed experimentally.

However, discrepancies exist between experimental and theoretical results:

(i) the model does not account for the observed decrease of $F_c$ with connectors length at low $V$,

(ii) no molecular weight dependence is experimentally noted for the relative increase of $F_c$ with $V$ (figure
\ref{fig8}b), whereas the model predicts a weaker velocity dependence for shorter connectors (figure \ref{fig9}).

We believe that these discrepancies have the following origin. In the model, the assumption is made that the chain is
pulled through but {\it stays} in the network, which is not the case experimentally since tethered chains are really
extracted from the elastomer. As discussed by Brown \cite{Brown}, this difference might have an impact on the absolute
value of $F_e$ since the extracted part of a connector, outside the network, is probably less confined than assumed
theoretically. Moreover,
the two lowest molecular weights of connectors used in our study are such that $Z/N_e$ is close to one, and are thus
weakly entangled with the elastomer, which may correspond to a situation at the limit of validity of the model and make
quantitative comparison difficult. Finally, at high velocity, we estimate from equation \ref{tau3} that the number of monomers in
the penetrated part is around 150, {\it i.e.} very close to $N_e$. Full disentanglement of the grafted chains at
high $V$ may thus account
for the drop of $F_c$.

\section{Conclusions}

We have perform an extensive experimental study of sliding friction at a rubber/brush interface. We have thus
been able, by means of a simple friction force measurement on a macroscopic contact, to access detailed
information on the molecular mechanisms at play during sliding at such an interface, and we have shown that:

(i) friction is very sensitive to the degree of heterogeneity of the surfaces: the presence of
defects in the brushes is evidenced by the appearance of unstable sliding (stick-slip), and variations of 10 to 20\%
on surface coverage could lead to fivefold variations on the sliding stress or the critical velocity at the instability
threshold,

(ii) in steady sliding, the interfacial response is clearly influenced by penetration of the grafted chains into
the rubber network. At low grafting density, frictional dissipation is governed by the chain pull-out mechanism, whereas
at high grafting density it seems to be dominated by the rheology of the thin entangled layer made of the elongated
extracted chains, confined out
of the elastomer,

(iii) our experimental data at very low grafting densities and the theoretical predictions by
Rubinstein {\it et al.} \cite{Rubinstein,Ajdari} are in satisfactory agreement as far as friction force level {\it and}
and velocity
dependence are concerned. This provides, to our knowledge, the first semi-quantitative evidence
that grafted chains
relaxation in the network during steady sliding
occurs by arm retraction, as previously
inferred
from stress relaxation experiments on network/brush \cite{Brown} or brush/brush interfaces \cite{Klein}.
Discrepancies observed on molecular weight dependence of the friction force suggest both
model refinements and experimental prospects. In particular, data obtained with different mesh sizes of the elastomer
(with a polymerization index between crosslink points $N<N_e$) would provide information on the contribution to friction
of the elastomer chains, and would allow to work in a less ambiguous regime where the polymerization
index of the grafted chain is much larger than that of the network.

\bibliographystyle{unsrt}

\begin{thebibliography}{99}
\bibitem{note} More precisely, for brushes made from the same wafer, if the reference brush ($\Sigma=0$)
has a low $\Sigma_{short}$, all the results obtained for $\Sigma >0$ are consistent with what has been shown in section
\ref{results} but systematically present high values of $\sigma_0$ and $V_c$. We conclude from this that all the bimodal brushes
made from the same wafer have a density $\Sigma_{short}$ close to that of the reference sample.
\end{thebibliography}



\end{document}